# Wavelength-scale stationary-wave integrated Fourier transform spectrometry

Etienne le Coarer[1,*], Sylvain Blaize[2,**], Pierre Benech[3], Ilan Stefanon[2], Alain Morand[3], Gilles Lérondel[2], Grégory Leblond[2], Pierre Kern[1], Jean Marc Fedeli[4], Pascal Royer[2]

[1]*Laboratoire d'AstrOphysique de Grenoble, Université Joseph Fourier, CNRS, BP 53, F38041 Grenoble cedex, FR*

[2]*Laboratoire de Nanotechnologie et d'Instrumentation Optique, ICD, CNRS (FRE2848), Université Technologique de Troyes, BP 2060, 10010 Troyes, FR*

[3]*Institut de Microélectronique d'Electromagnétisme et de Photonique INPG-UJF-CNRS (UMR5130), BP 257, 38016 Grenoble Cedex, FR*

[4]*CEA-LETI,Minatec 17 rue des Martyrs F38054 Grenoble cedex, FR*

* Contact author e-mail : Etienne.Le-Coarer@obs.ujf-grenoble.fr; sylvain.blaize@utt.fr

**Abstract:**

**Spectrometry is a general physical analysis approach to investigate light-matter interaction. However, the complex designs of existing spectrometers hinder them from simplification and miniaturisation which are vital for current intensive research in micro- and nanotechnology. Stationary-wave integrated Fourier transform spectrometry (SWIFTS) - an approach based on direct intensity detection of standing wave issued by either reflection (as the principle of colour photography by G. Lippmann) or counterpropagative interference phenomenon - is deemed able to overcome this drawback. Here, we present a SWIFTS based spectrometer relying on an original optical near-field detection method in which optical nanoprobes is used to directly sample the evanescent standing wave in the waveguide. Combined with integrated optics, we report an elegant way of reducing the volume of spectrometer to a few hundreds of cubic wavelengths. This is the first attempt of SWIFTS in achieving very small integrated 1D spectrometer suitable for applications where micro-spectrometers are essential.**

In 1891, at the "Académie des Sciences" in Paris, Gabriel Lippmann[1] presented a beautiful colour photograph of the sun's spectrum obtained with his new photographic plate. Later, in 1894[2], he published an article on how his plate was able to record colour information in the depth of photographic grainless gelatine and how the same plate after processing (development) could restore the original colour image merely through light reflection. He was thus the inventor of true interferential colour photography and received the Nobel Prize in 1908 for this breakthrough. Unfortunately, this principle was too complex to use. The method was abandoned a few years after its creation despite the considerable investment of Lumière brothers. At that time, one aspect of the Lippmann concept that was merely ignored relates to spectroscopic applications. Within the context of micro nanotechnology and miniaturisation of spectroscopic equipments that nowadays are the subjects of intensive research[3, 4, 5, 6], it becomes interesting to revisit the Lippmann concept. Actually early in 1933, Ives[7] proposed to use a photo electric device to probe stationary waves to make spectrometric measurements. More recently, in 1995, P. Connes[8] proposed to use arising new technology of detectors to make 3D Lippmann's based spectrometry. Following this idea, a first realisation of a very compact spectrometer based on Micro-Opto-Mechanical System (MOEMS) has been reported by Knipp et al.[9] in 2005 but with a very limited spectral resolution.

Based on the same concept, but taking advantage of photonics and near-field optics, we propose a new kind of Stationary Waves Integrated Fourier Transform Spectrometer (SWIFTS), in which direct sampling of evanescent standing waves is achieved using a collection of optical nanoprobes. The principle of SWIFTS is shown in Figure 1. Two configurations are proposed in which a direct near-field detection of confined standing waves is performed: the standing waves are issued either from a guided mode reflection as in G. Lippmann's principle of colour photography (figure 1a) or from two counterpropagative modes interference (figure 1b).

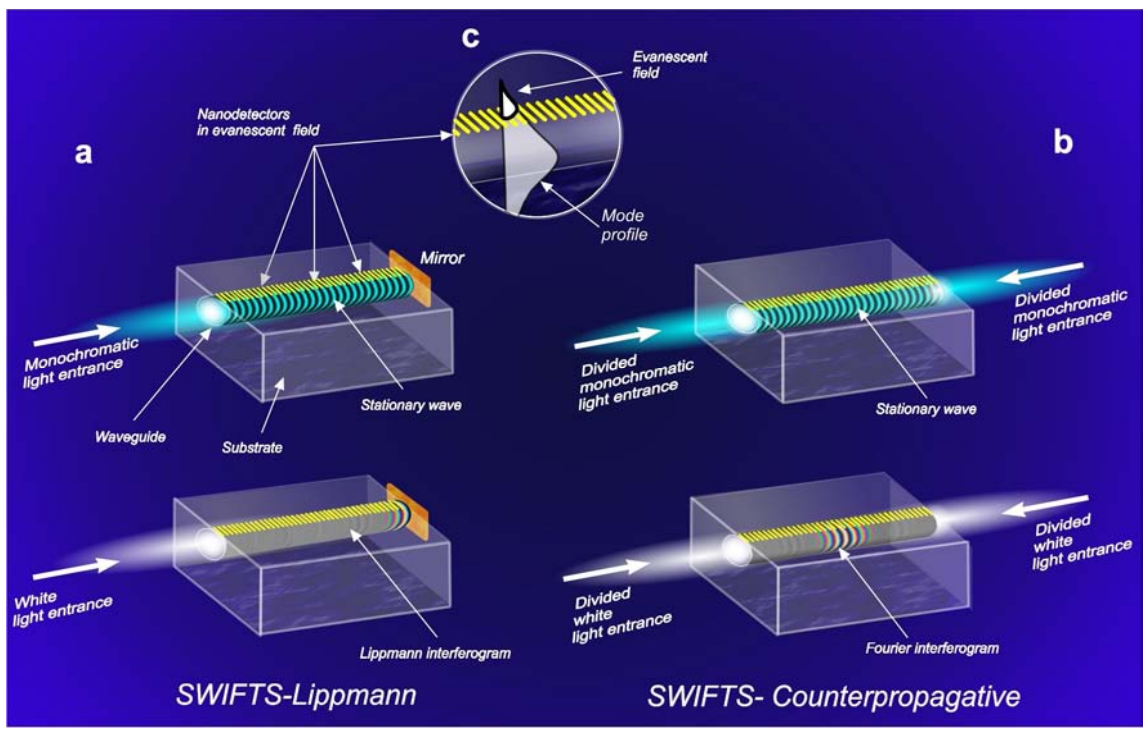

**Figure 1 : Stationary Wave Integrated Fourier Transform Spectrometry. a**, In the Lippmann onfiguration, the forward propagating wave coupled in the waveguide is reflected on the mirror, leading to a stationary wave. If light is polychromatic, the sum of the stationary waves forms a Lippmann interferogram. **b**, For the counterpropagative configuration, the light is divided upstream of the spectrometer. If the OPD between the two beams is null, the white fringe of the Fourier interferogram will be located at the waveguide centre. It is sensitive to the phase shift. **c**, Schematic of the near-field detection of the guided mode using a collection of nanodetctors.

In the first configuration, (figure 1a), light is coupled into a single mode waveguide terminated with a mirror. When reflected onto the mirror, waves become stationary. Miniature localized detectors are placed in the evanescent field (figure 1c) of the waveguide mode in order to extract as discussed further only a small fraction of the guided energy. This peripheral detection approach allows proper sampling of the standing wave using relatively small size detectors in comparison with the quarter wavelength of the guided light. The interferogram resulting from white light illumination (figure 1a) bears Lippmann's name. Unlike a classical Fourier interferogram, Lippmann's interferogram starts at the surface of the mirror with a black null optical path difference fringe such that only one side of the fringe packet is detected. In this way the whole energy is recovered using a minimum number of detectors. This principle thus acts like a spectrometer with simultaneous recorded Fourier transforms, i.e. no moveable part is required to record the information needed to restore the spectrum.

The second SWIFTS configuration (figure 1b) is based on the same near-field detection idea but instead of using a mirror, the light is injected from both sides of the waveguide. A similar configuration has been proposed by Labeyrie[10] for a data storage system which is an extension of holography[11,12] or of a Sagnac interferometer[13]. In this configuration, the resulting interferogram is a typical symmetric Fourier interferogram (Fig. 1b) which, in constrast with the mirror configuration, is sensitive to the optical path difference. This type of spectrometer could therefore also be used in metrology.

## RESULTS

### SWIFTS performes analysis

Before addressing fabrication details and issues, we would like first to discuss on the typical characteristics that one would expect from spectrometers based on the SWIFTS concept. These typical characteristics include spectral resolution, efficiency and bandwidth. As in classical Fourier spectrometry, the spectral resolution achieved by SWIFTS is given by the length of the detected interferogram: $R = \frac{\lambda}{\Delta\lambda} = \frac{2nL}{\lambda}$ where $n$ is the effective refraction index of the waveguide, $\lambda$ is the wavelength, $\Delta\lambda$ is the resolved wavelength by the spectrometer and $L$ is the waveguide length probed by the local detectors. In a single mode waveguide, the spectral resolution is only limited by the optical length ($nL$). Hence, SWIFTS spectrometer has no intrinsic limitation in resolution. For example, a SWIFTS spectrometer based on an optical waveguide including detectors placed over a length of 1cm ($10^7$nm), would allow for a spectral resolution of about R=40.000 (15 pm) at a 600nm wavelength.

P. Connes[8] has already note that an ideal Lippmann plate's efficiency can exceed 63%. Here, we theoretically and numerically demonstrate that an optimal efficiency of 74% can be obtained with SWIFTS assuming that each detector has ideal quantum efficiency and extracts exactly the same amount of energy $\eta_{loc} = 1/N$, where N is the total number of detectors (see § SWIFTS's internal efficiency in supplementary information).

Finally, the recoverable spectral range is ruled by two parameters: the sampling interval (i.e. the physical distance between two consecutive detectors) and the spectral range where the waveguide remains single mode. With a sampling interval satisfying the Nyquist's criterion, the recoverable spectral range is typically 400-500 nm for asymmetric waveguides at 1550nm, even more for symmetric waveguides and photonic crystal waveguides.

**Design and construction**

It is clear that to be fully efficient SWIFTS depends on the development of nanodetectors, which are not readily available as yet. For a very first demonstration, an intermediate solution based on light scattering was developed. The idea was to use nanometric scattering centres (see §Nearfield probe efficiency in supplementary information) deposited on the waveguide surface to extract the stationary field from the waveguide and thus make far field detection possible, using 2D planar detectors. Since we cannot use a detector array with a pitch equal to the quarter of the wavelength over a wide spectral domain, the Shannon's sampling criterion cannot be satisfied. However, if the spectral range is relatively small, the restoration of the spectrum still remains possible. This is known in Fourier spectrometry as the undersampling technique. For this first demonstration, the counterpropagative structure has been chosen (Fig. 1(b)). In addition to be sensitive to the optical path difference, this configuration avoids the technical problem of having a detector placed very close to the mirror.

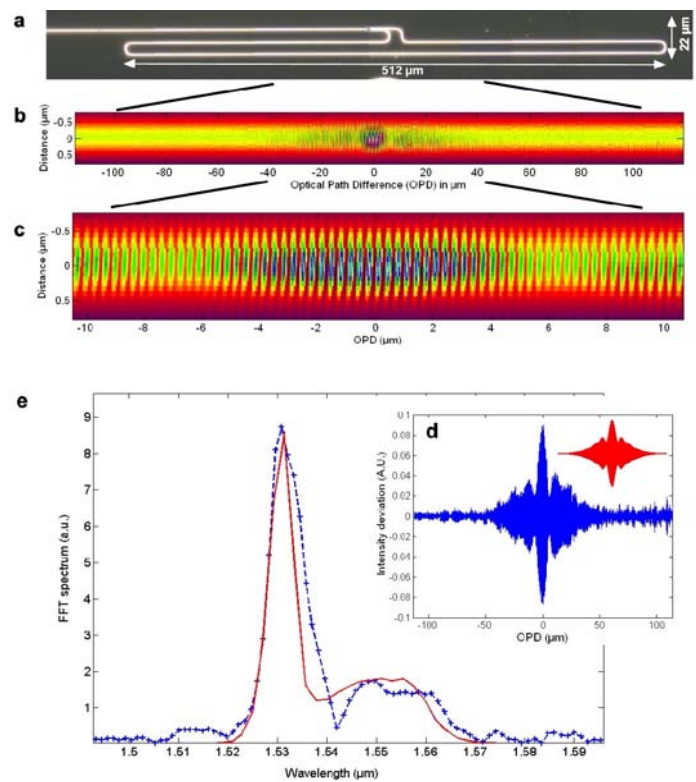

**Figure 2 s-SNOM observation of an interferogram in a waveguiding structure. a,** Top optical view of the interferometric component. The light of the ASE source is introduced into the guide and divided into two parts by an MMI, and propagates in opposite directions. The central fringe of the interferogram is equidistant from the MMI outputs. **b**, s-SNOM intensity image of the interferogram inside the component. **c,** 20 Â 1.5 mm2 enlarged image of the image in **b,** where the central fringes appear as fine vertical features. **d,** Measured (blue) and theoretical (red) interferogram profiles. **e**, Reconstructed spectrum in blue is compared to the manufacturer's ASE spectrum

In the following section, in order to validate the SWIFTS concept we will first focus on the probing of the interferogram inside a waveguide by means of scattering scanning near field optical microscopy. Secondly, the possibility of using nanometric scale dots to image the interferogram will be addressed and demonstrated. Hence we will show how a chain of nanowires can give a snapshot of the interferogram and thus allow recovery of the spectrum of an optical source.

**Probing the interferogram inside the waveguide**

The counterpropagative structure shown in figure 1(b) was fabricated using CMOS photonics technology. The silicon on insulator (SOI) substrate (200 mm wafer manufactured by SOITEC) used in this study is composed of a monocrystalline silicon layer with a typical thickness of 200 nm on top of a 1 μm thick buried oxide layer on a silicon substrate. After a silica hard mask deposition, the structure pattern is defined by means of a 193 nm deep UV lithography followed by hard mask etching and photoresist stripping. The pattern is transferred to the silicon layer by means of an HBr dry etching process. The entire waveguiding structure is then covered by a 700 nm thick cap layer of silica deposited by plasma-enhanced chemical vapour deposition. A chemical-mechanical polishing process is used to etch the capping layer until the planar structure has a uniform thickness of 100 nm.

An optical view of the full-size component, taken at the end of the process, is shown in figure 2(a). It mainly consists of a loop. The incoming light is introduced from the left side of the component into a straight single mode waveguide using a lensed fibre. To obtain single mode waveguides along a 1.5 to 1.6μm wavelength range, a waveguide width of 500 nm has been defined on the mask. A multimode interference splitter (MMI) is used to equally couple the light into the loop. The central fringe (also called white fringe) of the interferogram is found where the Optical Path Difference (OPD) is cancelled, i.e. into the loop waveguide at an equal distance from the MMI outputs.

In a preliminary study, the device was characterized using a scattering-type scanning near field optical microscope[14] (s-SNOM) which allows probing of the optical near-field of waveguides with nanometric resolution. A commercial Erbium-Amplified Spontaneous Emission (ASE) light source was used to couple the light into the waveguide. The broadband spectrum of this source in the telecom C-band centred around 1.55μm was sufficient to demonstrate the stationary phenomena which forms into the loop.

The optical image shown in Figure 2b was obtained using a numerical stitch of nine s-SNOM individual 50x50μm² scans in order to get a 50x220 μm² total scanned window centred on the white fringe position. As expected, the stationary nature of the guided field is revealed in the image by vertical lines (cf. figure 2c). Furthermore, the zero OPD can be localized where the standing wave visibility is maximal. The interferogram profile extracted from the s-SNOM image is shown in figure 2(e). Both measured (blue) and theoretical (red) stationary field intensity profiles are plotted in arbitrary units of intensity deviation for a better assessment. Since the counterropagative device acts as a Fourier transform spectrometer, the incoming light spectrum is recovered by computing the Discrete Fourier Transform of the stationary field profile. The result is shown in figure 2(d). The spectrum, i.e. the Fourier Transform of the interferogram is compared with the spectrum of the reference source (ASE) obtained with an optical spectrum analyser. The spectra are very similar. Some slight differences are attributed to stitching uncertainties in the individual s-SNOM scans. It is worth noting that only the total length of the scan limits the spectral resolution of the experiment. Presently, with a scan length of L=2×110 μm around the central fringe, one obtains a spectral resolution $R = \frac{\lambda}{\Delta\lambda} = \frac{2nL}{2\lambda} = 323$ leading to Δλ of 5nm.

**Towards the nanostructured device**

Although SNOM tips are ideal probes to sample the stationary field with nanometric resolution[15], SNOM is not compatible with the idea of compact integrated and monolithic optical devices. Therefore, in order to provide a parallel and motionless detection of the stationary field, we proposed the use of distributed nanometric light scattering defects embedded in the waveguide near-field. For that purpose, a comb of 79 gold nanowires, made by e-beam lithography and lift-off process, was deposited along the waveguide. Each nanowire oriented perpendicular to the waveguide has the following properties: first of all, it is sufficiently long (4μm) to interact with the total transversal field. Secondly, it is narrow enough (0.05 μm) to be considered as a point source scattering the light of the stationary field at a precise position. Finally, it is thin (0.05μm) enough to keep the volume small and thus only a small part of the field is scattered without disturbing the interferogram.

As shown by the AFM topography image (figure 3(a) and 3(b)), the crossing wires are arranged periodically with a pitch of 2.7 µm. Since the wires follow the initial rib Si waveguide topography, the wires look like a comb of equidistant pillars along the waveguide. Figure 3(c) is a far field image of the ASE stationary field sampled with the nanostructured waveguide. This image is the result of the scattering of 29 consecutive gold nanowires centred on the white fringe. Each light scattering defect is well resolved. Clearly, one can see that the detected signal is not perfectly symmetric from the zero OPD point. This is most probably due to a slight inconsequent misalignment between the central metallic wire and the white fringe maximum.

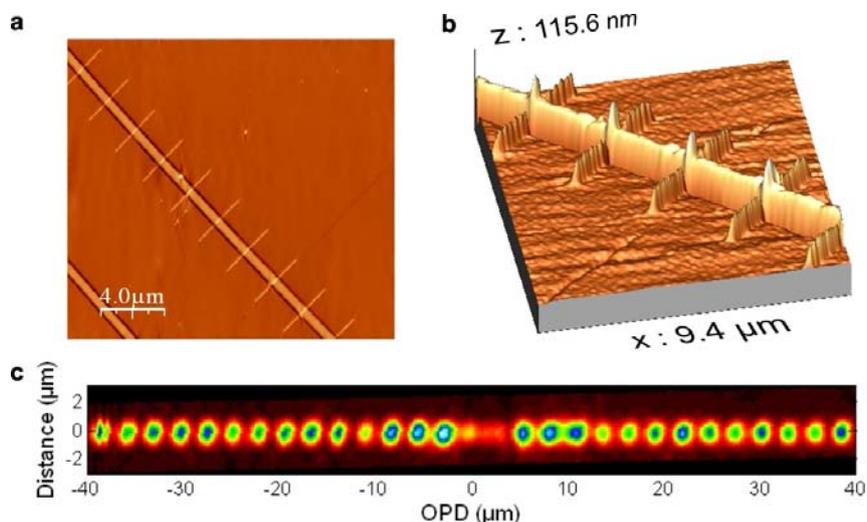

**Figure 3: nanowires structures**: **a)** AFM topography image of the nanostructured waveguide surface showing nine 4μmx50nm lithographic gold stripes crossing the silicon waveguide. **b)** Topographical 3D view **c)** Far field image of the ASE stationary field sampled by the 29 gold nanowires. The image is centred around the undersampled central fringe area

For calibration matters, the device was then characterized in the far field with a microscope objective, an InGaAs infrared camera and a single mode CW tuneable Er laser. Figure 4(a) shows a snapshot of the scattered light at a wavelength of 1510 nm. By computing the Fourier Transform of the longitudinal profile of the sampled interferogram, one obtains the laser light spectrum (Fig4(b)). This experiment gives an accurate calibration of the relation between the absolute wavelength and the measured spatial frequency. The process can be reiterated for the whole laser tuning range to obtain the curve in figure 4(c).

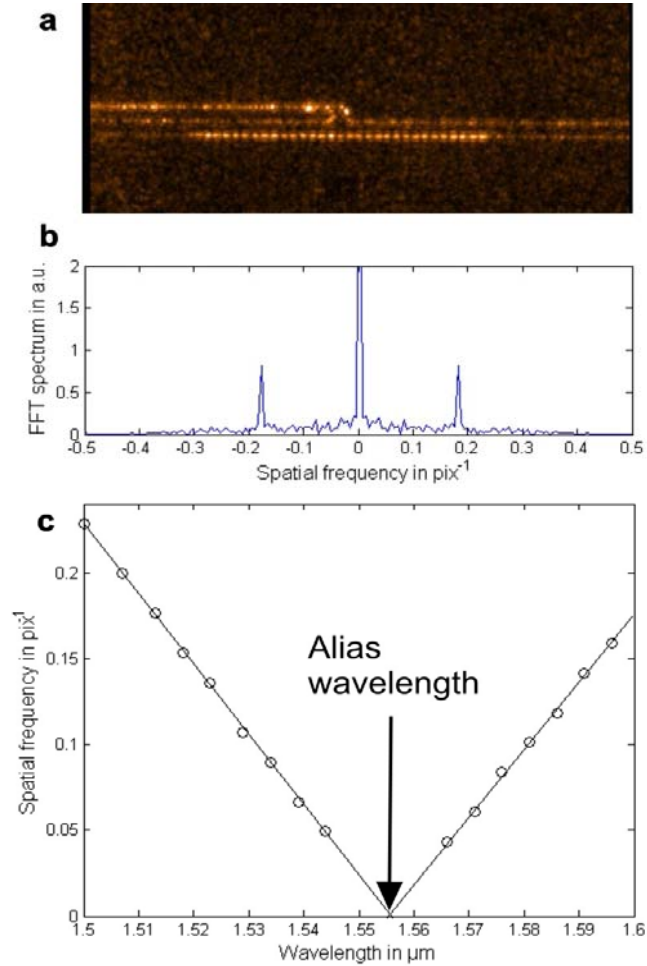

**Figure 4: monochromatic illumination**: **a)** image of the central part acquired with an infrared microscope. Tuneable laser fringes are seen as a chain of dots. **b)** Fourier transform of the recorded profile along the waveguide section on which gold nanowires have been deposited. The spatial frequency can be precisely deduced from the symmetric peaks. **c)** Plot of the spatial frequency versus the wavelength showing the aliasing effect. An alias wavelength at 1.55 µm corresponds to a 2.7 µm nanowire spacing with an effective refractive index of 2.301.

The wavelength limits of the recoverable spectrum can be found applying the band pass sampling theory[16]. For a stationary wave with spatial frequencies $f = \dfrac{2n_{eff}(\lambda)}{\lambda}$ lying in the range $[f_{min}, f_{max}]$, and a sampling frequency $m$ times lower ($m$ is an integer) than the Nyquist's frequency $f_N = 2f_{max}$, $dz$ is the distance between two nanowires. We obtain:

$$\begin{cases} \lambda_{min} = \dfrac{4n_{eff}(\lambda_{min})}{f_N} = \dfrac{4n_{eff}(\lambda_{min}).dz}{m} \\ \lambda_{max} = \dfrac{n_{eff}(\lambda_{max})}{n_{eff}(\lambda_{min})} \dfrac{m}{(m-1)} \lambda_{min} \end{cases}$$

As a consequence, any component of a sampled signal with a wavelength outside these limits, often referred to the aliasing window, is subject to folding. This is clearly observed in figure 4(c): wavelengths above the value 1.555 µm are folded into the aliasing window. Furthermore, the measurement of the alias wavelength leads to the value of the under sampling factor m=17. The sampling factor being a constant, this means that by measuring the position of the peak for each wavelength in the tuneable laser range, the dispersion relation of the waveguide $n_{eff}(\lambda)$ can be retrieved. As for polychromatic measurements a dispersion correction is needed, this procedure enables us to calibrate the SWIFTS spectrometer. The spectrum shown in Figure (4c) has been corrected for the effective refractive index dispersion.

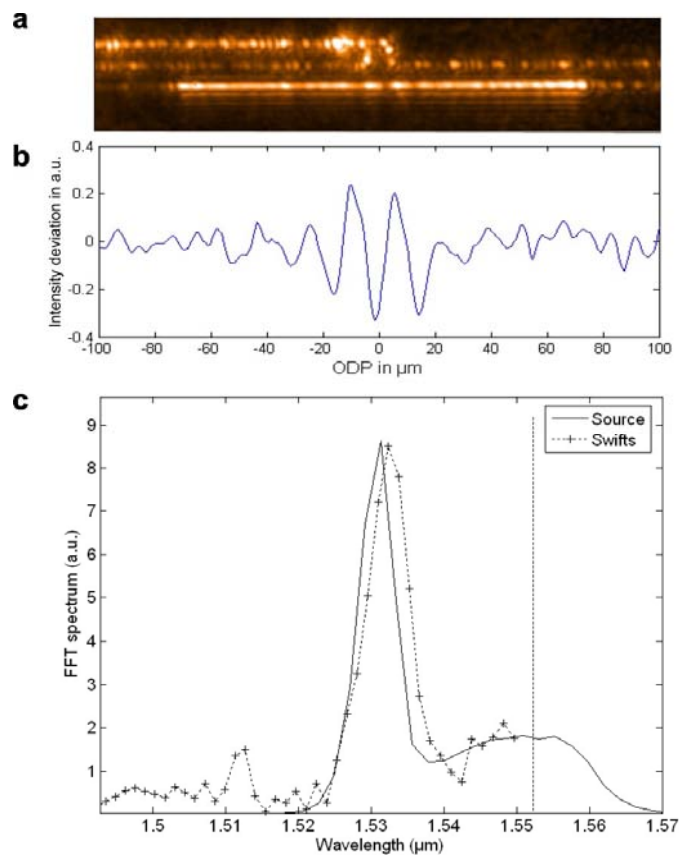

Figure 5: polychromatic illumination: **a)** microscopic image recorded with an infrared camera, the periodic dashed lines in fig. 4 are replaced by the Erbium fluorescence under-sampled interferogram. **b)** Extracted interferogram extending 200µm around the central fringe. **c)** Reconstructed (cross) spectrum, limited to long wavelengths by the aliasing windows, compared to the reference spectrum (solid line) recorded with a spectrum analyser.

Finally, the SWIFTS spectrometer was tested with an Er-ASE broadband source. The results are depicted in figure 5. The NIR camera snapshot is shown in figure 5(a). The stationary wave profile sampled by the gold wires is shown in figure 5(b). The spectrum recovered after calibration is plotted as a dotted line in figure 5(c) in contrast with the reference spectrum (solid line) obtained from a spectrum analyser. The Erbium fluorescence spectrum is well recovered up to the alias wavelength. This experimental result is very promising, especially if one takes into account the defects (merely surface scattering) inherent to a first realisation. However, a small unexpected bump is visible on the left side of the spectrum. It is probably due to a mismatch between the interferogram sampling interval and the CCD pitch and could be removed by an appropriate choice of the optical system magnification. Further work is required to fully understand this effect.

## DISCUSSION

The Lippmann concept of spectroscopy based on the Fourier Transform was never efficiently exploited until now. In this paper, a near field method to probe the interferogram within a waveguide has been proposed and demonstrated. The concept has been first validated using a scanning near field optical microscope. A first step towards fully integrated spectrometer has been made by local deposition of gold nanowires on the waveguide surface. Gold nanowires were 2.7 µm spaced in order to allow for far field probing using a simple CCD. With this structure it is clearly demonstrated that small gold wires are able to both scatter the light and sample the interferogram, allowing its detection.

The nanostructured waveguide we have developed can be seen as precursor of a new generation of spectrometers. This 1mm sized device has a spectral resolution of 4nm over a working spectral range of 96nm centred on 1500nm. At this stage a comparison with the State-of-the-Art integrated optical microspectrometers is untimely. However it is important to underline that the SWIFTS concept opens the route to very compact spectrometers with noteworthy features: high resolution, instantaneous measurement and no requirement for any moveable or tuneable parts. This has to be compared with classical Fourier Transform Spectrometry (FTS)[17,18] for which either the interferogram is sequentially recorded or statically projected onto a two dimensional detector array.

Further developments lie in using existing detectors arrays directly placed in the vicinity of the scattering centres. Fully integrated SWIFTS spectrometers will become manufacturable but with a limited efficiency. In future, the ultimate development will consist in replacing the gold nanowires by specially-designed detectors set every quarter of the standing wavelength. This will also enlarge the spectral bandwidth of the fully integrated spectrometer within the limitation of the single mode operation of the waveguides. It is worth mentioning here that a few of these nanoscale detectors already exist such as Superconducting Single Photon Detector (SSPD)[19] and carbon or silicon nanotubes.

Finally, it has to be stressed that while making spectroscopic instruments more robust and cheaper through the use of integrated technologies, SWIFTS is a general concept that paves the way for a number of applications and especially in optics where micro-spectrometers are essential. These applications would be space-borne spectrometry, metrology, endoscopy, gas and chemical sensors, colour photography and parallel spectral imaging.

**References**

1. G. Lippmann, La photographie des couleurs, *CRAS (Paris)*, **112**, 274-275 (1891)

2. G. Lippmann, Sur la théorie de la photographie des couleurs simples et composées, par la méthode interférentielle, *CRAS (Paris)*, **118**, 92-102 (1894)

3. Rommeluere, S. et al , Microspectrometer on a chip (MICROSPOC): first demonstration on a 320x240 LWIR HgCdTe focal plane array, Infrared *Technology and Applications XXX.~* Edited by Andresen, Bjorn F.; Fulop, Gabor F.~ Proceedings of the SPIE, **5406**, 170-177 (2004)

4. Wolffenbuttel, R.F., MEMS-based optical mini- and microspectrometers for the visible and infrared spectral range, *J. Micromech. Microeng.*, **15**, Issue 7, S145-S152 (2005).

5. J. Bland-Hawthorn, A. Horton, Instruments without optics: an integrated photonic spectrograph, Ground-based and Airborne Instrumentation for Astronomy. Edited by McLean, Ian S.; Iye, Masanori. Proceedings of the SPIE, **6269**, 62690N-1-14 (2006)

6. Froggatt, M. and Erdogan, T., All fiber wavemeter and Fourier-transform spectrometer, OptL., **24**, Issue 14, 942-944 (1999)

7. Ives, H.E., Standing light waves, repetition of an experiment by Wiener, using a photoelectric probe surface , *JOSA.*, **23**, 73-83 (1933)

8. Connes, P., Le Coarer, E., 3-D Spectroscopy: The Historical and Logical viewpoint *IAU Colloquium N 149*, Marseille, 22-25 Mars, 38-49 (1994)

9. Knipp, D., et al, Silicon-Based Micro-Fourier Spectrometer, *IEEE Trans. on Electron Devices* **52** (3), 419-426 (2005).

10. A. Labeyrie, J.P. Huignard, B. Loiseaux, Optical data storage in microfibers, *Opt. Lett.*, **23,** 4, 301-303,(1998)

11. Gabor, D., A New Microscopic Principle, *Nature*, **161**, 777-778, (1948)

12. Denisyuk, Y.N. , On the reproduction of the optical properties of an object by the wave field of its scattered radiation, *Opt. Spectrosc. (USSR)* **15**, 279-284 (1963)

13. Sagnac, G., Sur la preuve de la réalité de l'éther lumineux par l'expérience de l'interférographe tournant, *CRAS* (Paris) **157**, 708-710,1410-1413 (1913)

14. Stefanon, I. et al., Heterodyne detection of guided waves using a scattering-type optical near-field microscope , *Opt. Express*, **13**, n° 14, 5554-5564 (2005)

15. Bruyant, A. et al, Local complex reflectivity in optical waveguides" , *Phys. Rev. B*, 74, 075414-1 – 075414-16 (2006)

16. Lyons, R.G., Understanding Digital Signal Processing: Periodic Sampling, (Prentice Hall PTR, Upper Saddle River, NJ, USA, 2004).

17. Stroke G. W., Funkhouser, A.T., Fourier-transform spectroscopy using holographic imaging without computing and with stationary interferometers, *Phys. Lett.* **16**, 272-274 (1965).

18. Junttila, M.L., Kauppinen, J., Ikonen, E., Performance limits of stationary Fourier spectrometers, *J. Opt. Soc. Am.A*, **8**, 1457-1462 (1991).

19. Kadin, A.M., Johnson, M.W., Nonequilibrium photon-induced hotspot: A new mechanism for photodetection in ultrathin metallic films, *Appl. Phys. Lett.*, **69**, 3938-3940 (1996)

**Acknowledgments**

Authors warmly thank Sergei Kostcheev for the e-beam patterning of the scatters at the waveguide surface, Almas Chalabaev and Aurelien Bruyant for fruitful discussions, Gilles Duvert for the SWIFTS's acronym.

This work was partially supported by the CNES and the "Région Champagne Ardennes", and is part of the strategic research program on "Optical standing waves spectrometers and sensors" of the "Université de Technologie de Troyes (UTT)"

Correspondance and requests for materials should be addressed to ElC or SB

**Competing financial interestes**

The authors declare that they have no competing financial interests

Supplemental information for

# Wavelength-scale stationary-wave integrated Fourier transform spectrometry

Etienne le Coarer[1,*], Sylvain Blaize[2,**], Pierre Benech[3], Ilan Stephanon[2], Alain Morand[3], Gilles Lérondel[2], Grégory Leblond[2], Pierre Kern[1], Jean Marc Fedeli[4], Pascal Royer[2]

[1]*Laboratoire d'AstrOphysique de Grenoble, Université Joseph Fourier, CNRS, BP 53, F38041 Grenoble cedex, FR*

[2]*Laboratoire de Nanotechnologie et d'Instrumentation Optique, ICD, CNRS (FRE2848), Université Technologique de Troyes, BP 2060, 10010 Troyes, FR*

[3]*Institut de Microelectronique Electronique et de Photonique, UMR INPG-UJF-CNRS 5130 BP 257, 38 016 Grenoble Cedex, FR*

[4]*CEA-LETI,Minatec 17 rue des Martyrs F38054 Grenoble cedex, FR*

\* e-mail : Etienne.Le-Coarer@obs.ujf-grenoble.fr , sylvain.blaize@utt.fr ;

This supplemental gives the theoretical background necessary to address the internal efficiency of

SWIFTS-based spectrometers

**SWIFTS's internal efficiency – The continuous medium approach**

In order to give a theoretical description of the principle underlying SWIFTS, we introduce a continuous approach. We assume that every detector in the evanescent field takes the same ratio of energy and that the stationary wave is continuously attenuated. This can be modelled by a linear attenuation. We also suppose a monochromatic wave to simplify the demonstration.

Under these assumptions and considering only the SWIFTS-Lippmann configuration illustrated in fig 1.a), the interferogram intensity profile inside the waveguide is given by equation (1) which simply describes the two beams interferometry phenomena in an absorbing medium:

$$I(x, \lambda) = I_0 e^{-k(L-x)} + I_0 e^{-k(L+x)} + 2I_0 \sqrt{e^{-k(L-x)} \cdot e^{-k(L+x)}} \cos(\frac{4x\pi n}{\lambda} + \pi) \quad (1)$$

k is the linear attenuation factor in intensity.

$L$ is the length of the SWIFTS i.e. the length of the array of nanodetectors terminated by the mirror.

$n$ is the effective refractive index of the waveguide at the light wavelength $\lambda$.

and $\pi$ is the phase shift introduced by the reflection onto the mirror.

In order to emphasis on the useful signal which contains the spectral information, equation (1) can be dieivdvided into two parts: a pure interferometric term, $I_{int}$ and a slowly varying function $I_{svf}$ which should be minimized in order to reach the highest instrument efficiency:

$$I_{int}(x,\lambda) = 2I_0 e^{-kL}(1-\cos(\frac{4x\pi n}{\lambda}+\pi)) \quad (2)$$

$$I_{svf}(x,\lambda) = 2I_0 e^{-kL}(\cosh(kx)-1) \quad (3)$$

Along the waveguide, the linear detected intensity remains $k(I_{int} + I_{svf})$ and the total detected power becomes :

$$P = \int_0^L k(I_{int}(x,\lambda)+I_{svf}(x,\lambda)) \quad (4)$$

However, since only the pure interferometric term is valuable to recover the spectrum by a Fourier transform calculation, a signal processing consists first in subtracting the non useful part from the parenthesis. Combining (2) and (4), one obtained the total useful detected power remains:

$$P_{useful} = \int_0^L kI_{int}(x,\lambda) = \int_0^L 2kI_0 e^{-kL}.dx = 2kL \cdot I_0 e^{-kL} \quad (5)$$

where the contribution of the cosine term in equation (2) cancels with a fine adjustment of the device length L.

Finally, the SWIFTS efficiency, derived from equation (5) is given by

$$\eta = 2kLe^{-kL} \quad (6)$$

In conclusion, equation (6) shows that the SWIFTS efficiency is maximum for $kL=1$. This means that if N detectors are distributed along the waveguide, every detector has to probe 1/N of the local power. In this condition, 74% of the input coupled power contributes to the interferometric signal.

**Nearfield probe efficiency**

In the actual sensor concept, nanodetectors are replaced by nearfield optical probes consisting of gold nanometric wires. In order to describe more precisely this detection scheme, a 2D numerical model based on the Rigorous Coupled Wave Analysis (RCWA) combined with Perfectly Matched Layer (PML) has been developed. As described in J. Opt. Soc. Am. **19**, 335-338 (2002), the RCWA consists in periodically reproducing the index function of a region. This enables to solve the Helmholtz equation in this new space. The result of this equation is a Fourier decomposition. To limit the effect due to the periodicity, PML are added on each side of the region of interest (Suppl. figure 1.a). This method is applied to a 50x50nm² gold nanoparticle deposited on the top surface of a 200 nm thick Silicon planar waveguide (inset of Suppl. figure 1.a).

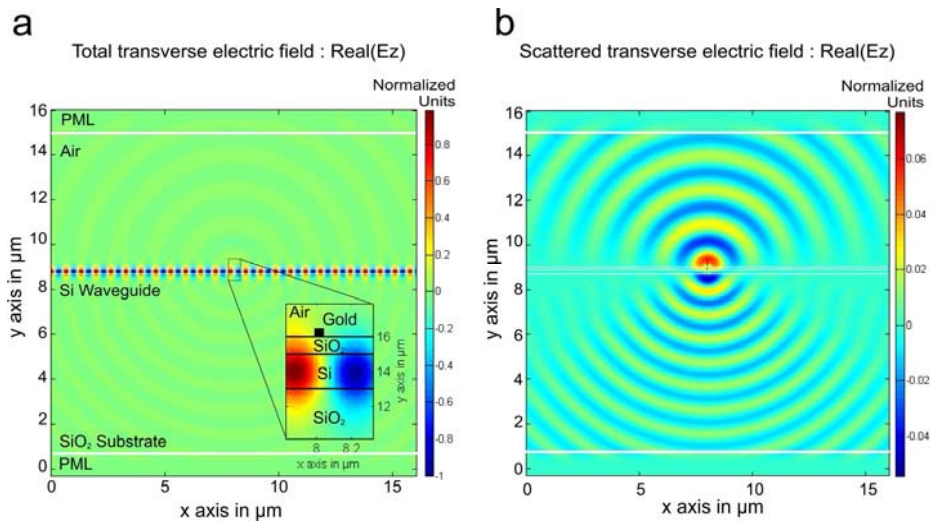

**Supplementary figure 1:** 2D numerical simulation of the detection principle underlying SWIFTS. **a)** Real part of the total transverse electric field. **b)** Real part of the scattered transverse electric field. In the actual sensor concept, the gold nanowire embedded in the evanescent field of the waveguide scatters up to 1.33% of the incoming light. These radiation modes can then be collected in far field by a CCD pixel. Calculations were performed using a RCWA based method.

The Suppl. figure 1.b illustrates the scattering of the guided electro-magnetic field in the region surrounding the waveguide. Despite the simplicity of this model compared to a realistic 3D description of the technological structure that slightly limits the accuracy of the analysis, this method is interesting to quickly evaluate the efficiency of the proposed design. From these numerical calculations performed at a wavelength of 1530 nm, the scattering power efficiency of the single metallic defect is estimated to be 1.33 % with a standard deviation less than 0.02% over the Er-ASE spectral band. The reflected power into the fundamental guided mode is less than 0.05% of the incident power and thus can be rightfully neglected. According to (6), it is straightforward to evaluate the optimal efficiency of the realized device. The calculated scattering efficiency of a single metallic defect of 1.33% is very similar to the optimal scattering efficiency $1/N$ here equal to 1.26% considering that $N = 79$ gold nanowires are embedded within the waveguide near field.